\documentclass[default]{sn-jnl}
\usepackage{amsmath,amssymb,amsfonts}
\usepackage{graphicx}
\usepackage{textcomp}
\usepackage{xcolor}
\usepackage{subfig}
\usepackage{caption}
\usepackage{multirow}
\usepackage[section]{placeins}
\usepackage{subfiles}

\jyear{2022}%

\theoremstyle{thmstyleone}%
\theoremstyle{thmstyletwo}%

\theoremstyle{thmstylethree}%

\begin{document}

\title[Exploiting Social Graph Networks for Emotion Prediction ]{ Exploiting Social Graph Networks for Emotion Prediction}

\author[1]{\fnm{Maryam} \sur{Khalid*}}\email{mk79@rice.edu}

\author[2]{\fnm{Akane} \sur{Sano}}\email{akane.sano@rice.edu}

\affil[1]{\orgdiv{Department of Electrical and Computer Engineering}, \orgname{Rice University}, \orgaddress{\street{6500 Main Street}, \city{Houston}, \postcode{77005}, \state{TX}, \country{USA}}}

\abstract{Emotion prediction plays an essential role in mental health and emotion- aware computing. The complex nature of emotion resulting from its dependency on a person’s physiological health, mental state, and his surroundings makes its prediction a challenging task. In this work, we utilize mobile sensing data to predict happiness and stress. In addition to a person’s physiological features, we also incorporate the environment’s impact through weather and social network. To this end, we leverage phone data to construct social networks and develop a machine learning architecture that aggregates information from multiple users of the graph network and integrates it with the temporal dynamics of data to predict emotion for all the users. The construction of social networks does not incur additional cost in terms of EMAs or data collection from users and doesn't raise privacy concerns. We propose an architecture that automates the integration of user's social network affect prediction, is capable of dealing with the dynamic distribution of real-life social networks, making it scalable to large-scale networks. Our extensive evaluation highlights the improvement provided by the integration of social networks. We further investigate the impact of graph topology on model's performance.} 


\maketitle
\newcommand\AS[1]{\textcolor{red}{AS:#1}}
\newcommand\MK[1]{\textcolor{blue}{MK:#1}}
\section{Introduction}\label{sec1}

Predicting emotion from passive sources can help regulate mental health, prevent breakdown/suicide, and make machines affect-intelligent. Mental health problems are impacting millions of people throughout the world with suicide being the third leading cause of death among young people \cite{suicide}. Mental illnesses not only impact an individual's work performance \cite{work} but also compromise the quality of life and relationships \cite{relationship}.  Emotion dysregulation is closely related to multiple mental health illnesses \cite{illness} that can be managed if emotions are tracked, and intervention is provided in time. Thus, emotion and well-being prediction using ubiquitous sensors and computing cannot only help humans regulate it but also make the machines more affect-intelligent. Machines and applications that incorporate user emotion in their operation can significantly improve the user experience.

Emotion  is a complex entity resulting from a human's current mental condition (internal dynamics) and external factors such as weather and social interactions.
This elaborate nature of emotion makes its prediction a challenging task. 
There is no device to directly measure emotion; however, there are several passive data sources that are indicative of a person's emotional state. Given the complex nature of emotion, one modality is not sufficient and data from multiple modalities need to be fused to obtain an accurate prediction.  Another important aspect of emotion is that an emotional state is caused by several temporally-correlated factors over time. Instantaneous measurements of modalities in most cases are not able to predict an emotional state with high accuracy. On the other hand, when viewed as a sequence, trends start to appear. %
Some data sources that are widely discussed in literature include video, speech and, wearable data \cite{speech}\cite{video}\cite{han1}, etc. Additional sources that are very easy to collect at high frequency without intrusion are physiological signals, weather, and information about the environment.

One important aspect of the environment is people. Previous studies show that people can transfer emotions to other people around them and this phenomenon is known as emotion contagion.
Formally, emotion contagion is defined as the ``phenomenon where the observed behavior of one individual leads to the reflexive production of the same behavior by others" \cite{contagion1}\cite{contagion2}. This can happen through face-to-face interaction, voice, text, or movements \cite{cont2} in an individual or group setting over time spans varying from seconds to weeks \cite{cont3}. Expanding on previous evidence suggesting the spread of emotion in a network, \cite{cont4} investigates whether happiness spreads from one person to another by leveraging graph networks. They utilize data collected over 20 years in Framingham Heart Study with 5124 participants and leverage the relationship and friendship information to construct graphs. After   conducting a regression statistical analysis on these graphs,  distinct clusters of happy and unhappy people are observed. Furthermore, their analysis showed that indirect ties till the depth of three and centrality in the network also impact future happiness. These works provide empirical evidence for the existence of emotion contagion but do not leverage it to predict emotion.

Multiple deep learning models have been developed to predict self-reported wellbeing scores using multimodal data such as physiological, behavioral, and social interaction data, many of which are summarized in \cite{survey1}. However, they do not account for user-to-user social interactions. The work in \cite{DA, han1, han2} predicts the next day's stress and happiness score from weather, physiological and behavioral data and accounts for the individual differences between users using personalized multi-task learning neural network models, gaussian process for domain adaptation, or fine-tuning neural network models such as convolutional neural network (CNN) and/or long-short term neural network (LSTM) to aggregate spatial and temporal aggregation of multimodal data. 
In addition to physiological data, these works also consider social interaction features such as the number of calls and sms made during a day. However, these models models do not combine information from multiple users to predict a users' wellbeing score. In this work, we account for the impact other people would have on a one's mood by integrating two aspects in the prediction model, user's own multi-modal data and information about other people that the user interacted with. To this end, we construct a prediction model which is complemented by an additional graph structure that allows the model to aggregate the information from multiple users.


There are multiple existing works that leverage graph architecture in emotion recognition and related applications, however, their objective is not to exploit contagion between people but extract complex relations between different modalities. Graph Neural Networks (GNN) are a popular architecture  \cite{gnnapp2}, \cite{gnnapp3}  \cite{gnnapp1} that have  recently gained a lot of attention in healthcare and mental health applications \cite{gnn0}\cite{gnn01}\cite{gnn2}. In all these works \cite{gnn0}\cite{gnn01}\cite{gnn1}\cite{gnn2}, graph representation is used to overcome the limitations of hand-crafted features with the objective of feature engineering. Our objective, on the other hand, is to exploit graph architecture to capture the role played by the  users' surroundings in their emotional state. Graphs created in previous related work are based on data from the \textit{same} user. However, graphs created in this work are composed of \textit{multiple} users.


\begin{figure}[t!]
    \centering
    \includegraphics[scale=0.42]{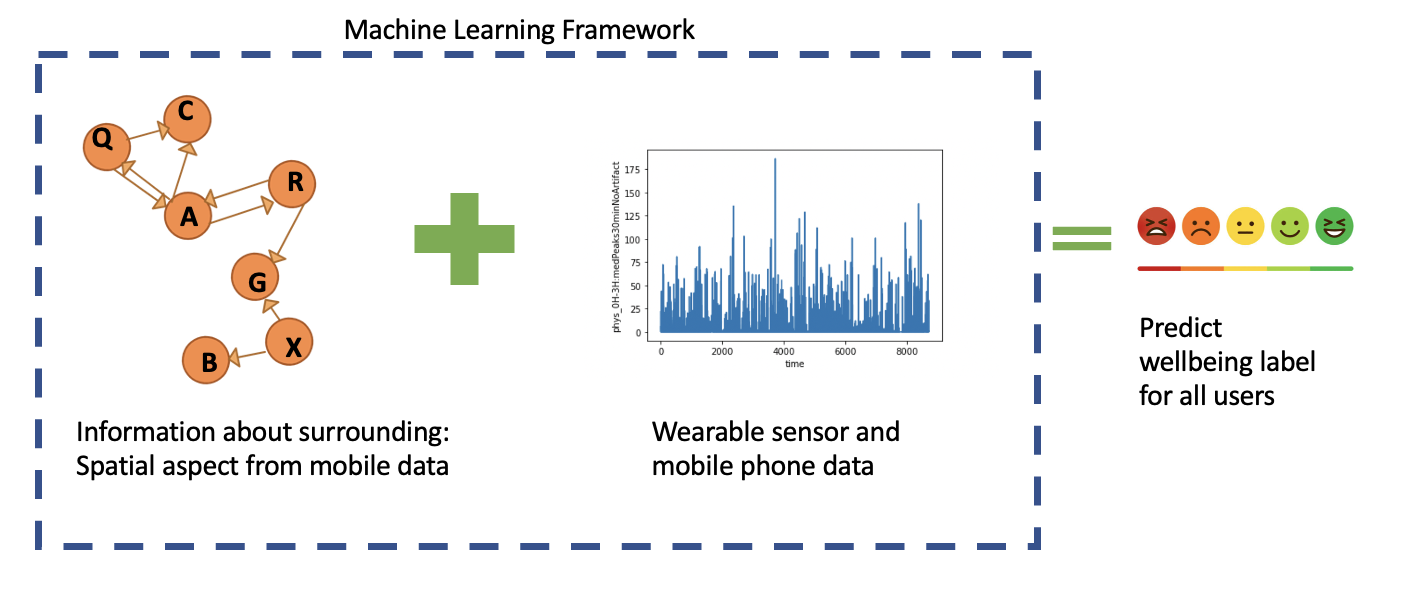}
    \caption{Core Concept: Graph networks incorporating user's contextual information about is integrated with time series wearable sensor and mobile phone data to predict user's emotional state. }
    \label{fig:core}
\end{figure}

In this work, we develop an emotion prediction model, 'GERIDSN: Graph-based Emotion Recognition with Integrated Dynamic Social Network' by integrating both temporal and spatial dynamics of physiological, behavioral, and social interaction information with graph convolutional neural networks and long short-term memory network as shown in Figure \ref{fig:core}. Inspired by the concept of emotion contagion, we exploit user's social network based on call and sms logs and design graphs where users act as nodes and call and sms interactions between them are quantified as connectivity links.
When information aggregation from other users in a participant's network is conducted in an automated way, a limitation is imposed on the size of the input graph: it should stay fixed. However, in real-world, network size can change dynamically. 
Graph convolutional network (GCN) in a supervised node classification problem cannot handle this dynamic user distribution. To overcome this problem, we present GEDD : Graph Extraction for Dynamic Distribution. Inspired by the information aggregation mechanism in GCN, our method leverages graph properties like connectivity and components to transform the set of varying size graphs into a set of graphs with fixed predetermined sizes. The proposed algorithm ensures that no users are discarded, and information is utilized to its full extent. Furthermore, it facilitates online learning with graphs where graph sizes are often changing.
We develop an architecture that facilitates integration of a user's social dynamics in his/her emotion prediction such that,

\begin{itemize}
   
    \item The sub-components of architecture are composed of GCN and LSTM layers that are differentiable and therefore can be easily trained using gradient descent methods. 
     \item  Extraction of graphs from existing phone data is automated.
    \item Can adapt to dynamic size of user's social network, that might change with time, allowing it to break extremely large networks into smaller networks in an efficient manner without any loss of information. With Internet-of-Things (IoT) networks and ubiquitous sensing, emotion recognition applications can be explored for macroscale networks and the proposed architecture can significantly facilitate these applications. 
\end{itemize}

We test our models using the data collected from over 200 college students who were socially connected as participants in 30-90 day seven cohort studies. The data collected in the study include (i) mobile phone data (call and sms logs, GPS, and screen usage), (ii) physiology (skin conductance (SC), skin temperature (ST), and 3-axis acceleration (AC)), (iii) surveys (daily emotions, drugs \& alcohol intake, sleep time, naps, exercise, academic and extracurricular activities) and (iv) weather (air pressure, humidity, wind speed, temperature, etc).

In an extensive experimental evaluation, the proposed model demonstrates improvement in emotion prediction accuracy resulting from the integration of user-to-user social interaction corroborating the findings on emotion contagion \cite{contagion2}\cite{contagion1}. The evaluation further explores the impact of network size that indicates the maximum neighbors taken into account by the model during prediction. The results indicate an initial improvement in accuracy with increasing network size until a plateau is reached after which diminishing returns are observed. This points toward the intricate dependency between graph topology, emotion, and prediction performance. To further gain insight into this conundrum, we conduct inferential statistical analyses between the influence of  users in the network, their emotional state and, the accuracy of the model. Our findings indicate a dependency of both, the emotional state and prediction error, on the eigenvalue centrality of the user.

\section{Results}\label{sec2}


\begin{figure}[t!]
    \centering
    \includegraphics[scale=0.65]{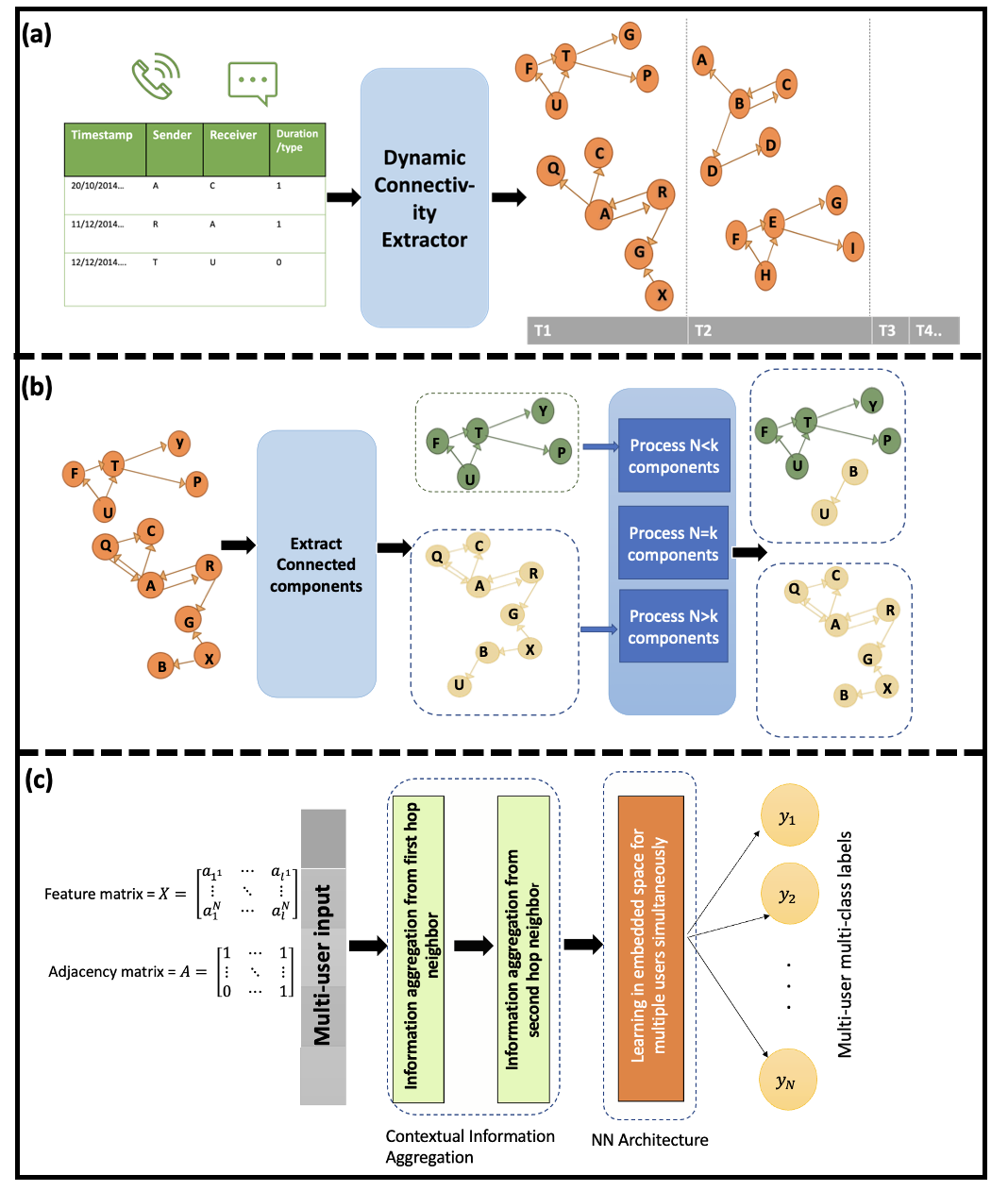}
    \caption{Multi-user Graph based learning framework (a) Graphs are constructed from call and sms logs for the entire study period  (b) GEDD converts varying size graphs to the desired size with minimal loss of information (c) Graphical structure is integrated with temporal sensor data in a GCN-LSTM architecture 
    }
    \label{fig:model}
\end{figure}

\subsection*{Graph Extraction}

The social interactions between participants are captured through a graph network which is composed of two main components: nodes and edges \cite{graph1}. We utilize the call and text message exchanges to establish weighted links between users as shown in Fig. \ref{fig:model}(a). We create two graphs: call graph $\mathcal{G}_c$ represented by adjacency matrix $\boldsymbol{A^c}$, and SMS graph $\mathcal{G}_s$ represented by $\boldsymbol{A^s}$ by aggregating call and sms information over a period T. 
 A challenging problem is posed by the varying number of participants in each interval $T$ as the proposed architecture  requires both features and graph network as input. Both these inputs \textit{fix} the size of the input layer. To overcome this problem, we develop an algorithm called Graph Extraction for Dynamic Distribution (GEDD) in Fig. \ref{fig:model}(b). The algorithm is fed with the extracted  graph $A$ and model input size $w$.  The algorithm breaks the $A$ into sub components and processes them to provide multiple subgraphs with $w$ nodes each.

\subsection*{Development of Learning Architecture}
We exploit the power of GCN to integrate a user's social interaction in the prediction process. The adjacency matrices processed through GEDD are integrated with features in the multi-layer GCN module in Fig.\ref{fig:model}(c). For temporal dynamics, the feature data for past $l$ days is fed as a time series to the LSTM-based module. The spatial and temporal dynamics are integrated through concatenation and batch normalization layers and finally fed to the dense layer. We label the final model GCN-LSTM for ease of notation. The overall model takes as input the temporal feature data and adjacency matrix for a given set of users and outputs the predicted emotion score of all the users.

\subsection*{Quantifying Improvement Provided by Network Integration}
We conduct an experimental performance evaluation to highlight the improvement in prediction provided by the aggregation of multiple users through a graph architecture. We hypothesize  that the integration of social interactions through a graph structure improves the prediction performance represented by the F1 score. To test this hypothesis, we compare with equivalent models that are similar in all other aspects except the ability to incorporate graph structure in the same experimental setting.
For comparison, we consider two other models:
\begin{itemize}
	\item \textbf{LSTM only:} In order to observe the improvement provided by graph integration, we also evaluate the model with LSTM layers only and no GCN. Like the proposed model, the LSTM layer is followed by multiple dense and dropout layers.
	
	\item \textbf{CONV-LSTM:} In recent work \cite{han2}, the convolutional neural network was utilized to aggregate information from multiple modalities. 
	For temporal dynamics, LSTM was used followed by dense and dropout layers.
\end{itemize}
We design the experiment to evaluate the performance and robustness of proposed scheme. We account for sensitivity to initialization and generalization in our experiment design. After preprocessing (details in the methods section), we train and test the models multiple times such that all three models are trained and tested on exactly the sample data samples within a trial.

The models predict the label for mood (stress and happiness) the next evening. The score for mood is categorized into three bins with class labels $0,1$ and $2$. Class 0 indicates no stress/happiness ($<33$), class 1 indicates moderate stress/happiness and class 2 indicates high stress/happiness($>66$). Since this is a multi-class problem, we utilize the F1 score as the performance metric. Moreover, since the problem is multi-class and the classes are imbalanced, we weigh all classes accordingly and therefore use  micro-average F1 score.

The performance results for empirical evaluation are reported in Table \ref{overall}. The F1 score and root mean square error (RMSE) for both stress and happiness indicate that the proposed GCN-LSTM model provides higher prediction accuracy and lower RMSE compared to the other two baselines. Please also note that the proposed model has a much lower variance in both RMSE and F1 compared to the baselines which are much more sensitive to the train/test split, initialization, and model hyperparameters. It is interesting to note that during stress prediction, CONV-LSTM baseline predicted values that were outliers and that lead to extremely large RMSE. However, when the F1 score is computed, the continuous scores are converted to categorical labels mitigating the huge impact of a few outliers on the overall metric. Additionally, the analysis of variance (ANOVA)\cite{anova} test is conducted to ensure that the difference between performance of all three models is statistically significant. It can be observed that all p-values reported in the description of Table \ref{overall} are below $0.05$ indicating that  prediction accuracy for the three models is statistically different.  

To further test our hypothesis about social networks boosting prediction performance, we utilize Tukey HSD post-hoc test for three cases. 
The results from the post-hoc analysis indicate that the population mean F1 score of the proposed model is higher than that of CONV-LSTM for both stress (Tukey HSD, p-value= $0.0003$) and happiness(Tukey HSD, p-value=$0.0054$). The proposed model also performs better than LSTM-only for both stress (Tukey HSD, p-value=$0$ ) and happiness(Tukey HSD, p-value=$0.000$). Furthermore, CONV-LSTM performs better than LSTM for both stress (Tukey HSD, p-value=$0.0023$ ) and happiness(Tukey HSD, p-value=$0.004$).

\vspace{5mm}
\begin{center}
	\begin{tabular}{|c|c|c|c|c|}
	
		\toprule
		&\multicolumn{2}{|c|}{\textbf{Stress}}
		&
		\multicolumn{2}{|c|}{\textbf{Happiness}} \\\cmidrule(r){2-4}\cmidrule(l){4-5}
		\textbf{Model} &F1$\pm$sd & RMSE $\pm$sd    & F1$\pm$sd & RMSE $\pm$sd     \\
		\hline
		GCN-LSTM&$0.69 \pm 0.02$ &$18.3 \pm 0.66$ &$0.72 \pm 0.02$&$17.3 \pm 0.8$\\
		\hline
		CONV-LSTM& $ 0.62\pm 0.03 $&$95\pm 88.5$ &$0.65 \pm 0.05$&$20.6 \pm 2.5$\\
		\hline
		LSTM & $0.57 \pm 0.03$ &$43.3 \pm 8.1 $& $0.57 \pm 0.04$  & $ 56.7 \pm 24.1 $\\
		\bottomrule
	\end{tabular}
		\captionof{table}{Performance comparison between different models. ANOVA test is conducted for statistical significance between different models. The test showed that all means reported in the table are statistically different for both stress (F1 p-value = $7.5\times 10^{-8}$, RMSE p-value =$2\times 10^{-2}$) and happiness (F1 p-value=$0.2\times 10^{-5}$, RMSE p-value=$1.6\times 10^{-5}$ )}\label{overall}
\end{center}

\subsection*{Impact of Graph Size and Sequence Length on Prediction Performance}
We further evaluate model performance for varying network sizes and sequence length. In first experiment, the input graph size is varied, and  the sequence length is fixed. The input graph size represents the amount of multi-user information that the model is exposed to for each sample. For a fair comparison between the models, the input size is kept the same across all models i.e if a graph of size $10$ is constructed and provided to the proposed method, then the features and labels of the same 10 users are provided to benchmark schemes as well. The train and test sets are kept fixed across all models.

\begin{figure}
    \centering
    \subfloat[][\centering Stress prediction]{\includegraphics[width=7.5cm]{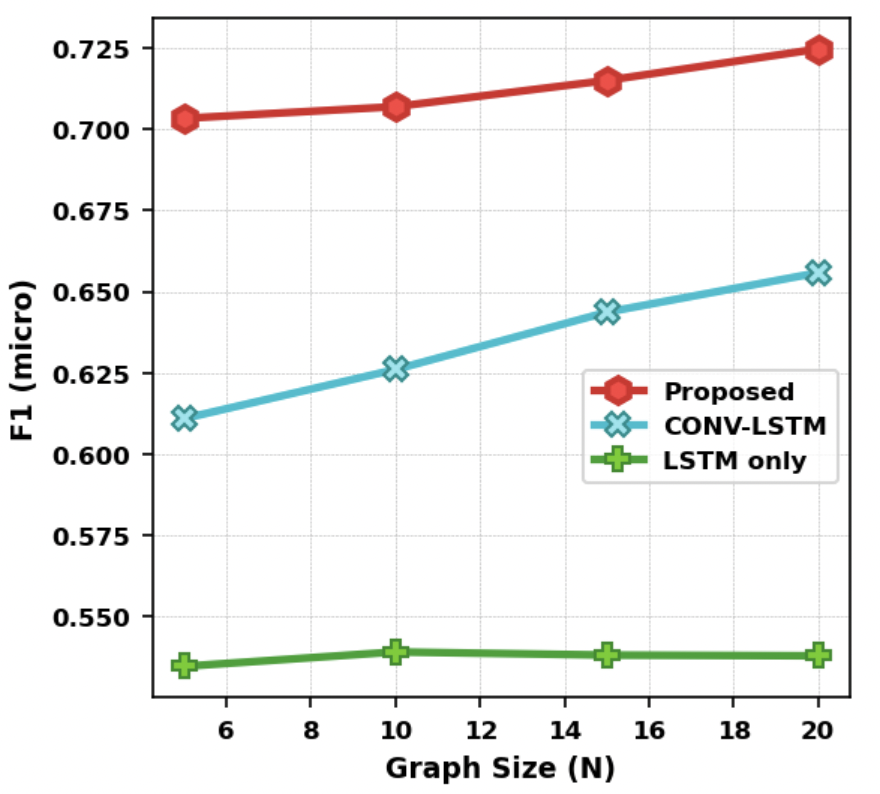} }%
    \qquad
    \subfloat[][\centering Happiness prediction]{\includegraphics[width=7.5cm]{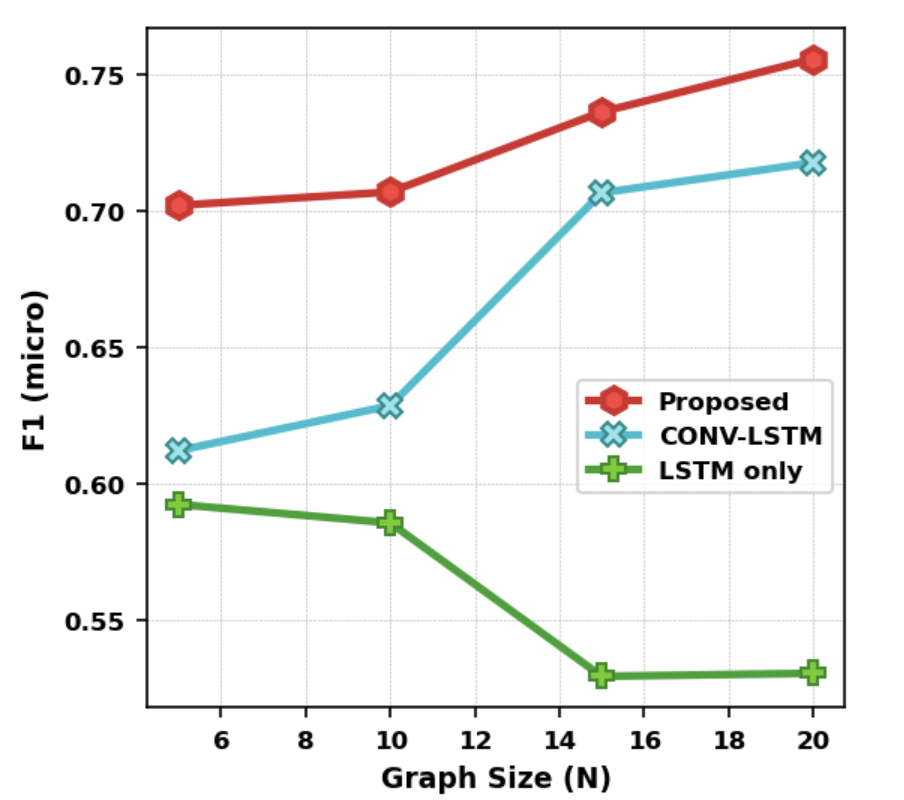} }%
    \caption{Impact of graph size on model prediction for fixed sequence length L=5.}
    \label{fig:graph_stress}%
\end{figure}
The results for the impact of graph size on  prediction accuracy are presented in Figure \ref{fig:graph_stress}. 
The plot (a) shows the  F1 score for stress prediction and the plot (b) shows performance for happiness prediction. Different lines represent the   proposed method GCN-LSTM and benchmark models. First, we observe that for proposed method and CONV-LSTM, an increase in graph size improves the performance. Even when the graph structure is not provided as in CONV-LSTM, a multi-user scenario is favorable for prediction compared to a single-user scenario. The slope of the proposed scheme is steeper in the beginning as we ascend towards larger graph sizes indicating higher gain in performance. However, after N=15, we see diminishing returns. This is because even though more nodes are added to the input graph, the edge density is becoming sparse and therefore no additional information aggregation is happening. Comparing the three models in both plots, we observe that the proposed method outperforms the other two schemes by a significant margin highlighting the role played by integration of social interactions in emotion prediction. Comparing the LSTM and CONV-LSTM, we observe that if aggregation between multiple users is not conducted in a systematic manner, it can be detrimental. Therefore, LSTM-only performs worse than CONV-LSTM alone. Additional results for RMSE are presented in supplementary information file in Fig.3:(S).

\begin{figure}[h!]
    \centering
    \subfloat[][\centering Smaller graph with 10 nodes]{\includegraphics[width=8 cm]{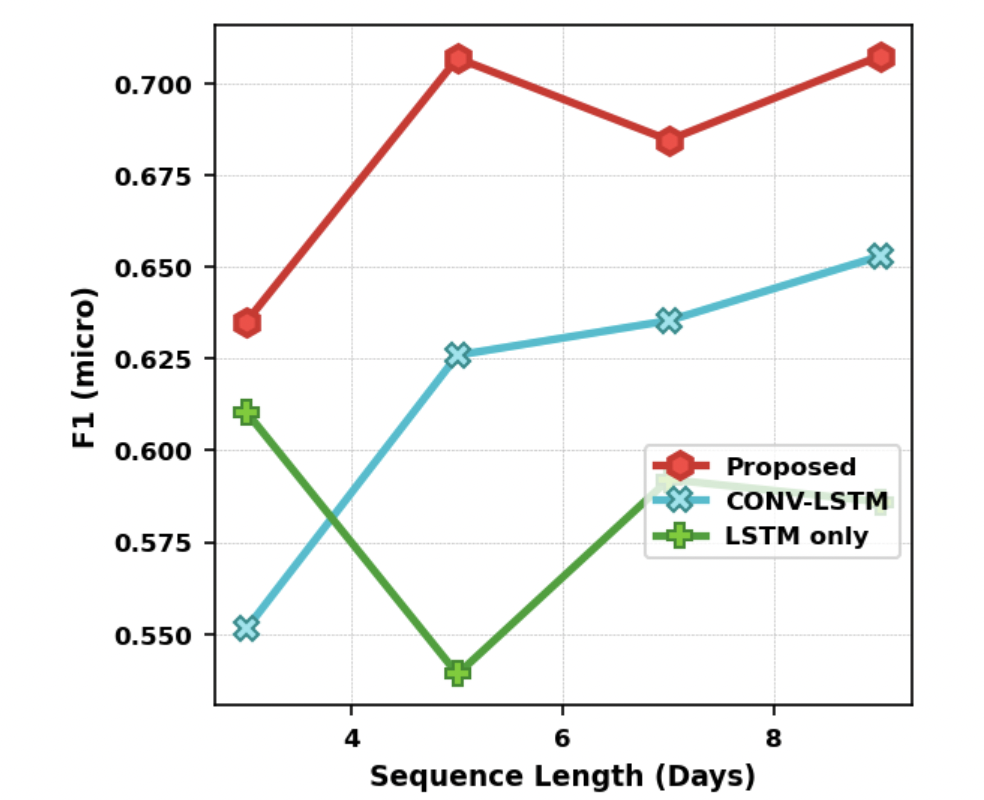} }%
    \quad
    \subfloat[][\centering Larger graph with 15 nodes]{\includegraphics[width=8cm]{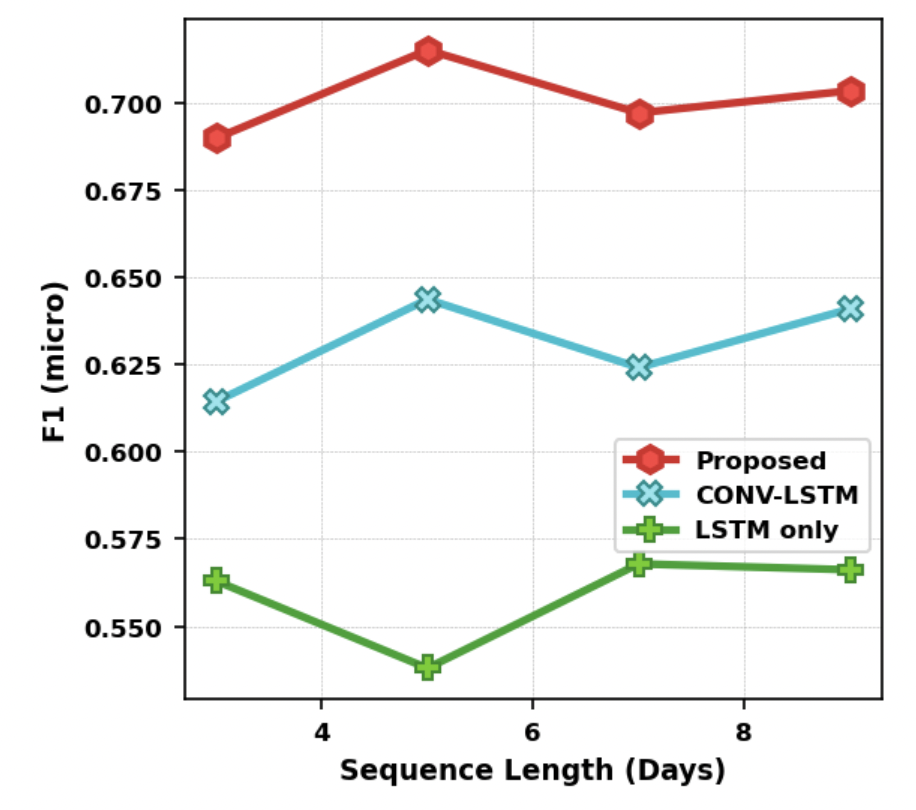} }%
    \caption{Impact of temporal memory on stress prediction}%
    \label{fig:seq_stress}%
\end{figure}

We further investigate the impact of temporal memory on performance and present the results in Fig. \ref{fig:seq_stress}. In this experiment, the graph size is kept fixed, and the time memory window is varied. Again, the number of input users is kept the same across all models and the same train/test data is provided for  a fair comparison.

Figure \ref{fig:seq_stress} (a) presents the results for small graph of size N=10 and (b) for larger graph N=15 for stress prediction. Looking at individual plots, we observe that the GCN-LSTM method performs significantly better than LSTM alone and CONV-LSTM. When temporal memory is low with just 3 days of past feature data, the performance is also poor. However, when 5 days of past data is provided, we see a boost in performance of all methods except LSTM-only. Further increasing the memory is detrimental as it introduces false dependency on much earlier points in the past which actually do not play a part in the current emotional state. If we compare (a) and (b) in Fig. \ref{fig:seq_stress}, L=3 is an interesting point. For this  short temporal memory the CONV-LSTM performs worse than LSTM which is a deviation from the average trend and hypothesis analysis conducted in Table \ref{overall}. However, when the input size is increased and multiple users are provided to the model at once, the model is able to \textit{make up} for the inability to capture temporal dynamics. Thus, we can conclude that the gap introduced by the unavailability of information provided by past data can be filled by utilizing the data from surrounding users. 
 
 The results for happiness prediction are presented in supplementary information file in Fig.2:(S). The performance improves with a larger graph size for all models. Similar to the trend observed in stress prediction, the performance improves when temporal memory is increased from 3 to 5 days and provides diminishing returns after that.


\subsection*{Network Characteristic Analysis}\label{stat_analysis}
To further gain insight into the  emotion transfer dynamics between people, we investigate the impact of network behavior on prediction accuracy. The graph network is characterized by multiple factors derived from its nodes' and edges' attributes. The  problem posed in this paper specifically focuses on contextual information aggregation from neighboring nodes and one key metric that can quantify this aggregation is  node centrality/importance. We utilize multiple representations of node centrality including eigenvalue centrality, pagerank, degree centrality, and closeness centrality.  The prediction accuracy of the model is determined by the average RMSE per user. To explore the relationship between node centrality and RMSE per user, the generalized estimating equation (GEE) is utilized. GEE provides a mechanism to estimate the parameters of a linear model while accounting for correlation among different observations of a group. We preferred GEE to other statistical linear models like mixed linear-effects models because we are interested in the overall relationship between network topology and model performance that represents the \textit{average} effect. Furthermore, GEE is robust to imprecise correlation structure. 

\begin{table}[]
\begin{tabular}{|l|ll|ll|}
\hline
\multirow{2}{*}{}                & \multicolumn{2}{l|}{RMSE for stress prediction} & \multicolumn{2}{l|}{RMSE for happiness prediction} \\ \cline{2-5} 
                                 & \multicolumn{1}{l|}{Coefficient}    & P-Value   & \multicolumn{1}{l|}{Coefficient}     & P\_Value    \\ \hline
Eigenvalue centrality            & \multicolumn{1}{l|}{3.5}            & 0.003     & \multicolumn{1}{l|}{3.4}             & 0.03        \\ \hline
Small Degree (D\textless{}4)     & \multicolumn{1}{l|}{-1.5}           & 0.02      & \multicolumn{1}{l|}{-1.6}            & 0.02        \\ \hline
Large Degree (D\textgreater{}4 ) & \multicolumn{1}{l|}{-1.9}           & 0.10      & \multicolumn{1}{l|}{-2.1}            & 0.11        \\ \hline
Closeness centrality            & \multicolumn{1}{l|}{1}            & 0.4     & \multicolumn{1}{l|}{0.8}             & 0.4       \\ \hline
Pagerank centrality            & \multicolumn{1}{l|}{-0.008}            & 0.6     & \multicolumn{1}{l|}{-0.01}             & 0.6        \\ \hline
\end{tabular}
\caption{Linear model between RMSE and node centrality metrics generated from GEE.}
\label{tab:gee}
\end{table}

The summarized results from the GEE analysis are shown in Table \ref{tab:gee}. The coefficients of the linear model between RMSE and node centrality metrics indicate the role played by that centrality in prediction and the p-value indicates the statistical significance of the learned coefficient. Please note that the results for closeness and pagerank centrality are not significant, so we can not use them for any inference. The degree centrality, which is the number of directly connected nodes, was categorized into two buckets: small degree and large degree with a threshold of 4 neighbors. For a small degree, we can observe that the coefficient is $-1.5$ indicating that the higher the degree, the lower the error because more information aggregation is happening. However, for large degree nodes, the results obtained are not statistically significant. This results from the tendency of the model to integrate information from irrelevant nodes during prediction. The same factors lead to a positive coefficient of $3.5$ for eigenvalue centrality. Eigenvalue centrality assigns higher scores to nodes that are close to influential nodes. Thus, during the prediction of such nodes, the model aggregates information from a large number of nodes that connect to its neighbors. Since these are not direct neighbors, there is a high chance of incorporating information from unimportant nodes leading to higher RMSE.

GEE is also utilized to fit a model between true stress/happiness scores and the node centrality metrics. The results for stress and happiness are shown in Table \ref{tab:true_stress} and supplementary information in Table 1: (S) respectively. It can be observed that eigenvalue centrality plays a role in both emotions. Higher eigenvalue results in lower average mood scores and higher standard deviation in those scores. This standard deviation also explains why nodes that are connected to influential nodes have higher RMSE.

\begin{table}[]
\begin{tabular}{|l|ll|ll|}
\hline
\multirow{2}{*}{}                & \multicolumn{2}{l|}{Average stress score} & \multicolumn{2}{l|}{Standard deviation in stress score} \\ \cline{2-5} 
                                 & \multicolumn{1}{l|}{Coefficient}    & P-Value   & \multicolumn{1}{l|}{Coefficient}     & P\_Value    \\ \hline
Eigenvalue centrality            & \multicolumn{1}{l|}{-7.1}            & 0.002     & \multicolumn{1}{l|}{3}             & 0.03        \\ \hline
Small Degree (D\textless{}4)     & \multicolumn{1}{l|}{0.96}           & 0.6      & \multicolumn{1}{l|}{-0.7}            & 0.23        \\ \hline
Large Degree (D\textgreater{}4 ) & \multicolumn{1}{l|}{-1.8}           & 0.4      & \multicolumn{1}{l|}{-1.3}            & 0.31        \\ \hline
Closeness Centrality & \multicolumn{1}{l|}{0.5}           & 0.9     & \multicolumn{1}{l|}{-0.6}            & 0.8        \\ 
\hline
Pagerank Centrality & \multicolumn{1}{l|}{0.01}           & 0.5      & \multicolumn{1}{l|}{-0.001}            & 0.5        \\ 
\hline
\end{tabular}
\caption{GEE results for model between true stress score and graph centrality metrics.}
\label{tab:true_stress}
\end{table}

\section{Discussion}

The deep learning architecture discussed in this work focuses on the prediction of multiple people’s emotion at the same time. Multi-user learning requires an informed mechanism to aggregate information from multiple users. It is intricate because there is correlation within features, correlation within users, and correlation between features and users. Convolution layers in CONV-LSTM explore this correlation in a more systematic way than LSTM and therefore performs better. This also explains why sometimes LSTM has overshooting RMSE in prediction results presented in supplementary section and thus a much higher standard deviation in error. The proposed model performs the best because it utilizes additional information about within user dependency. The integration of this dependency resulting in better prediction accuracy supports the existence of contagion and substantiates the impact one person’s emotion has on the other.

The way graphs are constructed to incorporate emotion contagion is also critical. Even though the performance evaluation indicated an overall improvement in prediction with the integration of social graphs, the network characteristic analysis revealed that higher eigenvalue centrality and a very large degree (greater than 4) negatively impact the performance. The phone data can indicate interaction with a large number of people however not all of them play a critical role in that person’s emotion. Similarly, when people are connected to influential people, the machine learning model would integrate information from the neighbors of those influential people many of whom might not have a significant impact. These findings suggest that more sophisticated graphs in an online learning framework like reinforcement learning should be explored in future research. Feedback, indicating which users are helpful for prediction during learning, would not only refine the graph and improve prediction accuracy but also yield deep insights into social contagion dynamics. With the availability of a huge amount of online data, the infrastructure proposed in this work, for constructing graphs and predicting emotions, is particularly useful for examining large-scale networks and therefore would be a resource for future research in macroscale social contagion.

The dataset used in this paper and other open-source emotion related datasets with wearable and mobile phone data are not designed to capture the social network of the participants entirely. There is a small likelihood that all friends and family of a participant are also a participant. This is a major limitation since it results in sparse graphs and a significant amount of the impact made by a user’s surroundings on his emotional state is not captured. The evaluation results presented in this paper show that graph-based architecture always performs better than the architecture that doesn’t account for it. However, there is still a large margin for improvement that graph-based architecture can provide if more dense graph networks are available.

Another limitation of this dataset is low temporal resolution of  social interaction. Since limited call and SMS logs are available, meaningful graphs can only be extracted when data spanning a time interval of a few days is considered. However, if more data is available for a larger number of participants, graphs can be extracted at hourly resolution and the prediction problem can also be solved at a higher temporal resolution.

When establishing a graph network between users from call and SMS logs, we noticed that participants were connected across cohorts as well. When taken into account, that provided us with a global graph network. However, since wearable and mobile phone data for those out-of-cohort connections were not available for the same time interval, we were not able to utilize them. In future works, algorithms for network sampling, estimation, and inference can be deployed to overcome this limitation and both global and local graph structures can be combined to improve the prediction performance.

One major challenge in emotion recognition problems is the collection of ground truth emotional state. There is no sensor to directly measure happiness or sadness. Therefore, we must rely on an individual’s judgment of their emotion which leads us to the second challenge associated with this problem: it is a user-centric study. Since the ground truth is collected through ecological momentary assessments (EMA), it is a tedious and expensive process. As graphs provide a global view of multiple connected users, they can be leveraged to identify
\textit{important} users whose data would be most beneficial for the overall prediction accuracy of the whole network. Once important users are identified, EMA data is collected from only them instead of the entire network.

The proposed work has significant utilization in real-world applications ranging from recommendation and regulation systems to web customization. A large amount of research works has focused on optimizing lifestyle by regulating activity, eating habits and sleep schedule\cite{app1}\cite{app2}. However, little attention has been paid to improving mental health by regulating social interactions and this work can play a pivotal role in such applications. This social interaction management can be taken a step further by developing apps that can customize the social media experience. Furthermore, the architecture proposed in this work can prove particularly useful in investigating emotion dynamics of people who spend a large amount of time as part of special environments such as healthcare workers, caregivers, rehabilitation counselors, etc.

\section{Methods}

\subsection*{Data Collection and Processing} 

We utilize a multimodal dataset 
collected in 2013-2017 from college students (age: 17-28, 146 male and 80 female) who were socially connected as participants. The study was conducted over multiple different time periods during each academic term for 4 years. Different students were recruited for 30-90 day studies in each academic term (N= 20-113 each in seven cohorts) . Four different types of data were collected in the study. After pre-processing of data, there are 314 features.

\begin{itemize}
	\item Mobile phone data : An app was installed on the participant’s phone that recorded the call logs, SMS logs, GPS, and screen usage along with timestamps. For phone features, statistics such as the mean, median, and frequency of these phone usage data were calculated for each time period (0-24H, 0-3H, 3-10H, 10-17H, 17-24H). Also, mobility features such as total daily distance, time spent on campus, and time with indoor/outdoor indications were calculated.
	
	\item Physiological data : From wearables, electrodermal activity (EDA) measured as skin conductance (SC), skin temperature (ST), and 3-axis acceleration (AC) were collected at 8 Hz. For each time period (0-24H, 0-3H, 3-10H, 10-17H, 17-24H), we calculated features about SC peaks and levels, ST, AC, and combinations of these physiological data streams.
	
	\item Surveys : Online surveys were filled by participants each morning and evening and contained information about drugs \& alcohol intake, sleep time, naps, exercise, and academic and extracurricular activities. All users filled out a survey indicating their calmness (stress) and happiness on a scale of 0-100 every day. We use these scores as ground truth in our problem.

	\item Weather: Data about weather conditions was extracted from Sky web API which was processed to extract air pressure, humidity, wind speed, temperature, information about sunlight and moon phase, and daily weather deviation from the rolling average.
\end{itemize}

The empirical distribution of both happiness and stress follows a Gaussian function. The mean and standard deviation for happiness score across all samples is  $61.8$ and   $23.8$. The mean and standard deviation for stress score across all samples is  $54.0$ and   $26.0$.
\\

\textbf{Ethics approval and consent to participate} The study protocols and informed consent procedure were approved by the Massachusetts Institute of Technology and Partners HealthCare Institutional Review Boards. The study was registered on clinicaltrials.gov (NCT02846077). All participants signed an informed consent form. All methods were performed in accordance with the relevant guidelines and regulations  complying  with the declaration of helsinki.


\subsection*{Models} The objective of this work is to predict the emotional state (stress and happiness) of a user based on multimodal data collected from wearable, mobile phone, and user-reported survey data. We focus on  well-being  prediction in terms of stress and happiness. 
The well-being metric for each day $n$ is represented as $y[n] \in [0,100] $.
Since the labels are not available at a frequency higher than 1 per day, we project the daily multimodal data to a compact representation such that for each feature we have one value per day by taking the mean and variance for different intervals of the day.  For ease of understanding, we represent this compact feature data for day $n$ in form of a vector $x[n]$. \\
For each $n$, the objective is to utilize the past $l$ days of information to predict $y[n]$,
\begin{equation}
	\hat{y}[n] = \underset{\theta}   {\mathrm{argmin}} \;\; \lVert y[n]- f( \boldsymbol{X_l^{n-1}} ,\theta )\lVert_2^2
\end{equation}
where $\theta$ represents the model parameters and  $l$ represents the time steps (memory) that model takes into account for prediction,
\begin{equation}\label{input}
	\boldsymbol{X_l^{n-1}}=[x[n-1] , x[n-2], ...,x[n-l-1]   ]
\end{equation}
Taking the mean and variance of different intervals of the day and representing each interval as a separate feature results in a huge loss of information. To reduce this loss, we deduce knowledge about how well users are connected from this data and utilize it to improve our model. We leverage graph networks to indicate the connectivity of participants (see more details about graph construction in the next subsection). We develop a weighted graph network $\mathcal{G}$ between a set of participants/users $\mathcal{V}$ whose connectivity or closeness is represented by a set of edges  $\mathcal{E}$, 
\begin{equation*}
	\mathcal{G=(V,E) }
\end{equation*} 

We can represent this graph as adjacency matrix $\boldsymbol{A}$ where the value at $i^{th}$ row and $j^{th}$ column is represented by $A_{ij}$,

\begin{equation}
	A_{ij} =
	\begin{cases}
		w_{ij} & \text{an edge exists $\mathcal{E}_{ij}$ exists from $\mathcal{V}_i$to $\mathcal{V}_j$ }\\
		0 & \text{otherwise}
	\end{cases}       
\end{equation}
And $w_{ij}$ represents the weight of edge.
The objective of the model is to predict $\hat{y}[n]$  with minimum error,
\begin{equation}\label{obj}
	\hat{y}[n] = \underset{\theta}{\mathrm{argmin}} \;\; \lVert ||y[n]-  f( \boldsymbol{X_l^{n-1}} ,\boldsymbol{A},\theta )\lVert_2^2
\end{equation}

\subsection*{Graph Extraction}

The social interactions between participants are captured through a graph network. A graph is composed of two main components: nodes and edges \cite{graph1}. Node is a vertex that is connected to other vertices through lines called edges.  In some problem settings such as social media networks, it is straightforward to establish a link e.g., when two users are friends, they are connected. However, if social media information is not collected and participants of a study do not provide information about whether they are friends with each other, creating edges is not straightforward. Even if users indicate friendship, as in the latter case, we need to define a graph  that is most helpful in achieving the prediction objective. Please note that one-time survey to identify whether users are friends with each other is not sufficient because they might be friends, but they do not interact very often because of different class schedules or circumstances. Instead, we utilize the call and text message exchanges to establish weighted links between users as shown in in \ref{fig:model}(a). We create two graphs: call graph $\mathcal{G}_c$ represented by adjacency matrix $\boldsymbol{A^c}$, and SMS graph $\mathcal{G}_s$ represented by $\boldsymbol{A^s}$. We design them based on phone data collected over an interval $[0,T]$. Representing an incoming call of duration $d$ from user $\mathcal{V}_i$ to $\mathcal{V}_j$ at time $t$ by $C_{ij}[t]$,
 \begin{equation*}
 	C_{ij}[t] = 
 	\begin{cases}
 		d & \text{$\mathcal{V}_i$ calls $\mathcal{V}_j$}\\
 		0 & \text{otherwise}
 	\end{cases}       
 \end{equation*}

\begin{equation}
	A^c_{ij} =  \sum_{t=0}^{T}C_{ij}[t]
\end{equation}

For text messages, we consider two types of incoming messages: normal SMS with a text message body (Class 1 message), and flash SMS with no message body (Class 0 message). Denoting an SMS from user $\mathcal{V}_i$ to $\mathcal{V}_j$ at time $t$ by $S_{ij}[t]$,
 \begin{equation*}
	S_{ij}[t] = 
	\begin{cases}
		w_1 & \text{$\mathcal{V}_i$ sends $\mathcal{V}_j$ a Class 1 message}\\
		w_2 & \text{$\mathcal{V}_i$ sends $\mathcal{V}_j$ a Class 0 message}\\
		0 & \text{otherwise}
	\end{cases}       
\end{equation*}
Prioritizing Class 1 messages because of their stronger interaction $w_1 > w_2$, we construct SMS graph,
\begin{equation}
	A^s_{ij} =  \sum_{t=0}^{T}S_{ij}[t]
\end{equation}
For each node in the adjacency matrix, there is associated feature data $X$ and label data $y$.
The time interval $T$ is equal to each cohort's study interval which is approximately equal to a month.

\subsection*{Contextual Aggregation from Multiple Users} 

When considering a multi-user scenario, there are two further sources of information that need to be exploited to make predictions about emotional state. The first source is the individual's  feature vector $X$, computed from the mobile phone, wearable, and survey data. The second is the relationship between multiple users i.e. the structural properties of the graph. When considering feature data for each user independently, neural networks can learn the underlying model and provide predictions. Further improvement can be made by exploiting local structure by deploying convolutional neural networks that use kernels/filters to extract complex features from a grid-like structure\cite{MLbok}. However, they cannot be utilized in this problem because they cannot operate on a graph-like structure. The flattened adjacency matrix cannot be utilized as input to these models because the neural network is not permutation invariant i.e it depends on the ordering of nodes in the adjacency matrix. This problem is addressed by GCN proposed in \cite{GCN}.

Inspired by spectral convolutions on graphs, GCN provides a layer-wise linear propagation rule that allows a neural network to learn from graphs. Spectral graph convolution is the convolution of any signal $x$ with a filter $g$, where the filter $g$ is derived from the graph\cite{specconv}. In order to compute spectral convolution, we need the laplacian and degree matrix of a graph. The degree matrix $D$ is a diagonal matrix containing degrees of the nodes on the diagonal where degree of a node is sum of incident edges. 

For more details on GCN layer, please refer to supplementary information section 1 and \cite{GCN}\cite{wavelet}. The forward propagation at layer $f^{th}$ of multi-layer graph convolutional neural network is represented by $H^f$ , 
\begin{equation}\label{main}
	H^f = \sigma (   \tilde{D}^{-1/2} \tilde{A} \tilde{D}^{-1/2} H^{(f-1)}\Theta^f  )
\end{equation}
where $H^f$ represents $f^{th}$ layer of GCN, $\tilde{A}$ represents the graph adjacency matrix,  $\Theta^f$ are the weights of $f^{th}$ layer and $\sigma$ is the activation. The input to the first layer is the node feature matrix $X$ defined in eq(\ref{input}). 
This equation is very similar to that of a dense layer of a conventional neural network except  that the degree and adjacency matrix representing the graph aggregate the inputs from previous layers based on user connectivity. 

To predict well-being labels using this model, labels for each user/node are used to compute the cross-entropy loss, and the model parameters are learned through forward and back propagation. The training process and parameter tuning are similar to the conventional neural networks. The only difference is that in addition to feature matrix $X$, graph adjacency matrix $\tilde{A}$ is also computed through call and SMS interaction data in our project and used as input to the model.

The graphs extracted for different cohorts vary in size in the range 20-50 nodes. The minimum and maximum number of directly connected neighbors of a given participant are $0$ and $12$ respectively. The average number of direct neighbors across all cohorts is $1.2$ with a standard deviation of $2.2$.

\begin{figure}[t!]
	\includegraphics[scale=0.45]{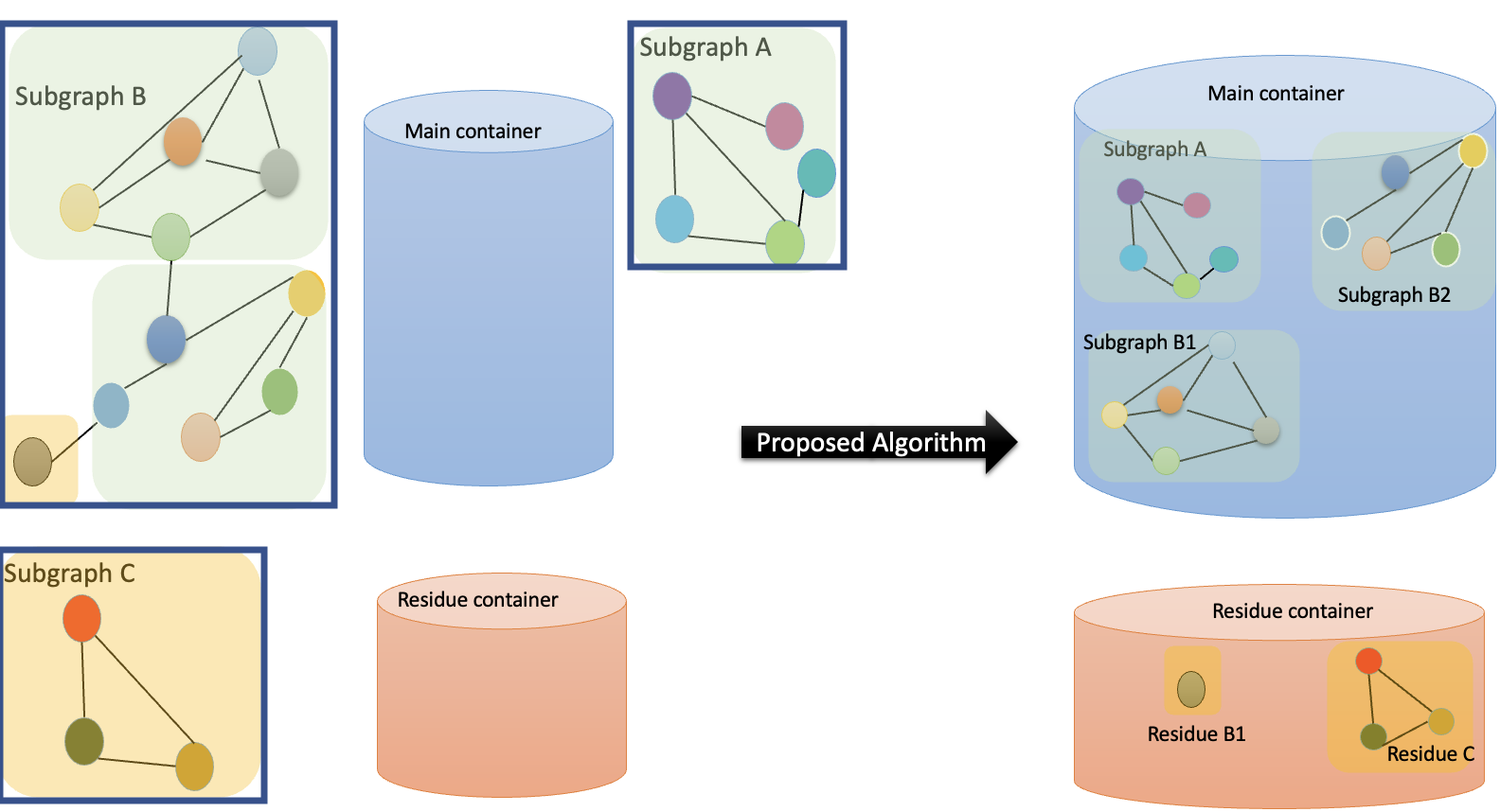}
	\centering
	\caption{Framework to extract subgraphs of size $w$ for dynamic user data. The proposed algorithm GEDD, first converts large components into smaller subgraphs of size $w$ with a residue. The optimal and adjusted graphs are added to the main container and the residue subgraphs of size $<w$  to the residual container. At the end the residual graphs are processed to obtain required size $w$ graphs   }\label{fig:gedd}
\end{figure}

\subsection*{Learning Temporal Dynamics} 
When utilizing multi-modal data for well-being prediction, it is important to realize that most well-being indicators are not impulsive and independent entities. Several factors lead to a certain state. It is very likely, that evening health conditions could not be explained from data collected in the morning but from the temporal dynamics of the same features for the past few days. To integrate these dynamics into the model, an element of ’memory’ is required that can utilize contextual information to predict future emotional states.

For the emotion prediction problem, we extract the sequential information in features to predict the well-being label $y$. For a given user, let $x[n]\in \mathbb{R}^N$ and $y[n]$ represent the  feature vector and well-being score for day $n$ respectively.  If $p\in  \mathbb{Z}^+$ is the length of sequence, then for we create an $N\times p$ sequence matrix,
\begin{equation}
	S_n^p =\big[x[n-p]] ,x[n-p+1],...x[n]\big]
\end{equation}
This sequence matrix serves as an input feature matrix for which we predict the future well-being label $y[n+l]$. Here $l\in  \mathbb{Z}^+$ represents how far in the future we want to make the prediction. The tuples $(S_n^p,y[n+l])$ are used to train an LSTM network in a supervised fashion with the cross-entropy loss function. The model weights are learned through conventional forward and backward pass over the model with gradient descent. 

\subsection*{Graph Extraction for Dynamic Distribution of Users}\label{GEDD} The data were collected from different sets of socially connected participants in each cohort. Each cohort has a different number of participants who enrolled in the study on slightly different study start and end dates.

The variation in the number of participants per cohort is challenging because the proposed GCN-LSTM model requires both features and graph network as input. Both these inputs \textit{fix} the size of the input layer. When the number of users changes, the size of the graph changes from the model’s pre-determined input size. Conventional methods used in image classification such as padding, or truncation are not feasible because graphs are different from images and have a more non-uniform structure. Truncation of such a structure would result in a huge loss of information. When a node is discarded from the graph, not only the graph structure information is lost but also the user’s feature matrix comprising of a large amount of physiological, mobile, and survey data.  Furthermore, padding the adjacency matrix with zero values would result in singular degree matrices in eq. \eqref{main} forcing the model to output undefined values.

To overcome this problem of varying user sizes, we propose an algorithm called Graph Extraction for Dynamic Distribution (GEDD). It is a connected component-based method that converts large dynamic graphs into a set of small graphs of size equal to the model's input size $w$. Our method is inspired by the working principle of the graph convolution network presented in eq. \eqref{main}. GCN exploits graph structure by combining a node's feature vector with its neighbor's information.  In the first layer, information from one-hop neighbors is aggregated. In the second layer, information from 2-hop neighbors is integrated. As we go deep down the network, the knowledge from farther away neighbors is aggregated.  Therefore, for label prediction of a given node $w_i$, a node $w_j, j\neq i$ would only contribute if it is connected directly or through a multi-hop connection to $w_i$, and nodes that are not connected are irrelevant for $w_i$'s prediction.

We exploit this concept for extracting graphs of size $w$ by utilizing connected components. A connected component of a graph is a subgraph in which each node is connected to another through a path\cite{net1}. For a graph with $N$ nodes $	\mathcal{G_N=(V,E) }$, there are $p$ connected components with $1\leq p\leq N$. When $p=1$, all the nodes in 	$\mathcal{G_N }$ are connected, and when $p=N$, all nodes are disconnected and have 0 degree. The breakdown of graph in connected components will result in subgraphs of varying sizes. Let $C_i$ 
represent the $i^{th}$ connected component,
\begin{equation*}
	C_i = \{\mathcal{V}^j,\mathcal{E}^j\} \;\; i={1,2,..p}\;j={1,2,,...N}
\end{equation*}
and $\|C_i\|=q_i, 1\leq q_i \leq N$ represent the size of the component. 
First, the components are divided into two containers, Main container $\mathbb{M}$ and residue container $\mathbb{R}$, based on their size as shown in Figure \ref{fig:gedd}. The former   will contain subgraphs of size $w$ and the residual will contain graphs of size $r< w$. This leads to three scenarios,
\begin{itemize}
	\item when $q_i=w$, add $C_i$ to  $\mathbb{M}$
	\item when $q_i<w$, add $C_i$ to  $\mathbb{R}$
	\item when $q_i>w$, break $C_i$ into $j=\lceil \frac{q_i}{w}  \rceil$ subgraphs $C_i^b$ where $b={1,2,..j}$. The large component is broken  such that,
	\begin{equation*}
		C_i^b =
		\begin{cases}
			w & \text{ $b={1,2,..,j-1}$}\\
			q_i \bmod w & \text{$b=j$}
		\end{cases}       
	\end{equation*}
	The subgraphs that satisfy first condition in above equation $\{C_i^1,C_i^2,...,C_i^{j-1}\}$ are added to $\mathbb{M}$  and $C_i^j$ to $\mathbb{R}$.
\end{itemize}
Once the components are divided between two containers, the main container is ready to be fed to the model. For the residue container with all subgraphs smaller than $w$, we concatenate multiple  subgraphs to create size $w$ subgraphs. There is still some residue left at the end when the \textit{total} number of nodes in $\mathbb{R}$ are less than $w$. For this last set, we use repetition of nodes to create a final size $w$ subgraph. 

\subsection*{Experiment Design} 

We design the experiment to evaluate the performance and robustness of proposed scheme and account for sensitivity to initialization and generalization in this process.

\textbf{Preprocessing.} The pre-processing of data involves two main steps, dealing with missing data and standardization. First, features with more than $\lambda/2$ missing values are removed, where $\lambda$ is the total number of samples. For remaining features, the filling data is filled in two main steps. In the first step, the missing entries for a given user are filled using its own data through k-nearest neighbor imputation. This method identifies k neighbors for a datapoint and replaced the missing value with the mean value of those neighbors. In the second step, we consider all users together and repeat the k-nearest neighbor imputation for the entire dataset. Finally, we detect outliers using z-score statistic and remove them. Z-score is computed by subtracting from the datapoint its mean   and dividing by the standard deviation.

As mentioned in the dataset description section above, the feature matrix is composed of multiple modalities that come from very different distributions. In order to account for the difference between features' scale and spread, we standardize the data i.e we transform the distribution of data such that it has a 0 mean and unit standard deviation. To achieve this transformation, first, the mean of data is subtracted from it and then this zero-mean data is divided by its standard deviation.

\textbf{Performance Metrics.}
The model predicts the label for mood (stress or happiness). Since this is a multi-class problem, we utilize the F1 score as the performance metric. Moreover, since the problem is multi-class and the classes are imbalanced, we weigh all classes accordingly and therefore use  a micro-average F1 score. To compute the F1, first we calculate micro-average precision $P$ and recall $R$,
\begin{equation}
	P =  \frac{\sum_{p=1}^{3} TP_p  } {\sum_{p=1}^{3}  TP_p + FP_p   } 
\end{equation}
\begin{equation}
	R =  \frac{\sum_{p=1}^{3} TP_p  } {\sum_{p=1}^{3}  TP_p + FN_p   } 
\end{equation}
where  $TP_p$, $FP_p$ and $FN_p$ represent true positives, false positives, and false negatives for class $p$ respectively and $p$ denotes the class ID. Finally, the micro-average F1 score is computed as follows,
\begin{equation}
	F1 = 2* \frac{P*R}{P+R}
\end{equation}

\textbf{Learning Pipeline.}
For learning the model parameters, $50\%$ of data is used for training, $10\%$ for validation, and the remaining for testing. In order to remove the impact of sensitivity of the model to parameter initialization, we repeat the training and testing procedure ten times and report average results along with the standard deviation across trials. Furthermore, the size of the dataset is of the order of a few thousands which is small in comparison to feature space and model complexity. Therefore test-train split can impact the performance. To account for this,  we split the data into test and train sets randomly and  repeat the procedure ten times. 

We utilize ADAM optimizer with the same learning rate for all three models. During the training procedure, validation loss is monitored, and the model comes to an early stop if validation loss is not changing by more than $\Delta=10^{-5}$ for more than 50 iterations. This helps save computational time and cost.

\textbf{Modeling Graph Behavior.} Graph network is not only indicative of  user clusters but also the complex interconnectivity. The notion of connectivity has many aspects which are captured by different types of centrality metrics defined in literature\cite{net1}. We hypothesize that the more central a user is in the network, the more information aggregation would happen in the prediction model and that would impact the model prediction performance. The centrality metrics used to quantify the influence of a person  are listed below,

\begin{itemize}
	
	\item \textit{Degree Centrality} is  a direct representation of how many directly connected neighbors a node has. If the graph is represented by $N\times N$ adjacency matrix $A$, then the degree centrality $C_d$ of  a node $v$ is calculated as,
	\begin{equation}
		C_d(v) = \sum_{u=1}^{N}\mathcal{I}(A_{uv})
	\end{equation}
	where, 
	\begin{equation}
		\mathcal{I}(x) = \begin{cases}
			1 & \text{if  $x > 0$}\\
			0 & \text{x = 0}
		\end{cases}   
	\end{equation}
	
	Degree centrality assigns higher importance to nodes that have a large number of neighbors. However, it does not account for the cascade effect resulting from the fact that a node can also be important if it is connected to influential nodes. 
	
	\item \textit{Closeness Centrality} represents the importance as to how close a node is to other nodes in terms of geodesic distance. To compute the closeness centrality $C_c$ of a node $v$, the shortest distance to all other nodes is computed,
	\begin{equation}
		C_c(v) = \sum_{u\in \mathcal{V}}\frac{N}{ d(u,v)}
	\end{equation}
	where $d(u,v)$ is the shortest path from node $v$ to $u$. This is  computed using Dijkstra's algorithm\cite{dijstra}.

	\item \textit{Eigenvalue centrality} quantifies the influence of a node in a network by measuring the node's closeness to influential parts of a network. It combines the degree of a node with the degree of its neighbors. For a graph $\mathcal{G}$ with adjacency matrix $A$, the eigenvalue centrality $C_e$ of a node $v$ is calculated by\cite{net2},
	\begin{equation}\label{eigen}
		C_e (v)=  \alpha \sum_{u,v\in \mathcal{E}}C_e(u) 
	\end{equation}
	where $C_e$ and $1/\alpha$ are the eigenvector and corresponding eigenvalue of $A$ respectively,
	\begin{equation}
		AC_e = \alpha^{-1}C_e
	\end{equation}
	Please note that solving the eigenvalue problem for large graphs is expensive. In this scenario, the power iterations method is used to compute the  eigenvalue and the corresponding eigenvector for a graph  with $N$ nodes and $\mathcal{O}(N_v^2)$ complexity,
	\begin{equation}
		C_e(z+1) = \frac{AC_e(z)}{|| AC_e(z)||}
	\end{equation}
	where $z$ represent the iteration index.
	
	\item  \textit{Pagerank Centrality }assigns high importance to nodes who are connected to important nodes, or  if they are  linked by a lot of other nodes who themselves have small outgoing connections. Thus, it incorporates both, the importance of neighboring nodes and the number of incoming edges. It was proposed by \cite{pr} to retrieve relevant pages from the web in response to a query. 
	To calculate pagerank centrality $C_p$ for adjacency matrix $A$ , first, the in-degree $d^{in}$ and out-degree $d^{out}$ is calculated for node $v$,
	\begin{equation*}
		d^{in}(v) = \sum_{u=1}^{N}A_{uv} ,   \;\;\;\;\;\;\;\;\;\;\;  d^{out}(v) = \sum_{u=1}^{N}A_{vu} 
	\end{equation*}
	\begin{equation}
		C_p(v ) = \gamma \sum_{u=1}^{N}\frac{A_{uv}}{d_u^{out}}C_p(u) +\frac{1-\gamma}{N}
	\end{equation}
	where $N$ is the number of nodes if the graph and $\gamma$ is a constant damping factor. For nodes, with no outgoing links, the algorithm would get stuck and therefore, such nodes are known as sinking nodes. To avoid this problem, damping factor is introduced that prevents the algorithm from terminating when ending in such sinking nodes. 
\end{itemize}

\subsection*{Outcome Metrics for Statistical Analysis} 
The overall evaluation metric for the model is RMSE. However, since the test samples are chosen randomly and the model is trained and evaluated multiple times, multiple predictions for different users with different graph characteristics are obtained. Furthermore, there are participants whose connectivity dynamics changed over time and so do the resulting centrality scores. While defining the outcome metric, it is important to ensure that the model distinguishes between the performances of the model for a participant in different graph topologies i.e. a user can have different prediction accuracy when its centrality changes. To achieve this, we define RMSE per user and compute multiple RMSE scores for the same user for different graph topologies. At the end of multi-trial evaluation, we filter out the predictions for user $i$ when it was in a topology $j$ and create the column vector  $y^p_{i,j}$ with true labels $y^t_{i,j}$ and compute the RMSE $R_i^j$,
\begin{equation}
	R_i^j = \sqrt{\frac{\sum_{d=1}^{D}(y^p_{i,j}[d]- y^t_{i,j}[d] )^2}{D}}
\end{equation}

where $y[d] $ denotes the $d^{th}$ entry of the vector $y$ and $D$ is the length of vector $y^p_{i,j}$.

\subsection*{Statistical Analysis} 

After defining the independent variable representative of graph characteristics and dependent evaluation criteria, we investigate the relationship between them. We fit a linear marginal model between the two variables in a clustered data analysis setting while accounting for within-cluster and between-cluster variations.

Fitting the same model to all the participants and treating the whole population as one cluster would result in over-simplification of the underlying complex model. While the emotional state of a person is affected by his surroundings and social interactions, the magnitude, and type of this effect varies from person to person. We capture this individual customization through personality traits and create personality clusters.

We fit a generalized linear model between the expected value or RMSE and graph-characteristic covariate vector. In order to obtain population-level estimates of model parameters, we utilize GEE\cite{gee}. Assuming $P$ clusters with $n_p $ observation in $p^{th}$ cluster, where $p=1,2...P$. Let the RMSE for $p^{th}$ cluster and $q^{th}$ observation be represented by $R_{pq}$ and corresponding $w\times 1$ covariate vector $X_{pq}$. The response vector for cluster $p$ is denoted by $R_p=[R_{p1},R_{p1},...,R_{pn_p}]$ with expected value $\mu_p=[ \mu_{p1}, \mu_{p2},...,\mu_{pn_p}  ]$,
\begin{equation}
	R_p = \mu_p + \epsilon
\end{equation}
where $\epsilon$ represents the random error term.
We fit a linear model between covariates and the expected value of response vector   $\mu_p$ \cite{gee2},

\begin{equation}
	g(\mu_p) = X_p*\beta
\end{equation}
where $g(.)$  is known as the link function that depends on the distribution of the response variable and $\beta$ is $w\times 1$ vector containing regression coefficients that need to be estimated. At this point, there are three key design parameters: identification of groups, the probability distribution of response variable, and the working correlation structure of the RMSE variable.

\subsection*{Group Identification} Participants fill a one-time Big Five personality\cite{big5} survey at the start of the study. The questionnaire response is then processed to obtain a score for the five personality dimensions, extraversion, agreeableness, conscientiousness, openness, and neuroticism on a scale of 1 to 100.
In addition to personality traits, we also incorporate gender information. 

Based on these six features, the participants are clustered into groups. We use hierarchical clustering that sequentially partitions data and creates a hierarchy of clusters. In order to identify the optimal number of clusters, we build a dendogram of all observations \cite{clus1}. Then we cut the tree diagram horizontally such that it captures more than $70\%$ of the data and count the number of clusters above the cutting line.

For these optimal number of clusters, we apply Agglomerative clustering to the data. This clustering method works its way bottom-up starting by treating each object as a cluster, then merging pairs of clusters until the desired number of clusters is reached. For merging clusters, the similarity between sets of observations is quantified by computing a distance metric called linkage between observations across the two pairs of clusters. We utilize ward linkage as it gives the most balanced clusters. We identify 11 optimal clusters with minimum and maximum euclidean distances equal to  46 and 136 respectively. The cluster sizes vary between 20 and 40 points.

\textbf{Distribution of Response Variable.} The link function maps the expected value of the response variable to the linear regression of covariates and is derived from the distribution followed by the response variable. Therefore, we plot the histogram of the  RMSE as shown in supplemetary information in Figure.1:(S). It can be seen that it closely follows a Gaussian distribution. Since the response variable is normal, no transformation is needed and we use the identity link function\cite{gee3}. 

\textbf{Correlation Structure.} The correlation structure of the response variable accounts for the correlation between different participants within a cluster. For cluster $p$ with $n_p$ observations, the working-correlation  $\Sigma_p$ is an  $n_p \times n_p$ matrix with diagonal entries equal to subject variance and cross-diagonal entries representing the inter-subject correlation. A summary of commonly used working correlation structures is provided in \cite{gee2}. We conducted multiple experiments with commonly used structures listed in \cite{gee2} and observed that the fitted model has the highest confidence in estimated parameters for Autoregressive structure defined by,
\begin{equation}
	\Sigma_p  (R_{pi},R_{p(i+m)}) = \alpha^m \;\;\; for \;\;\; m=0,1,2,...,n_p-j 
\end{equation}
where the parameter $\alpha$ is estimated from current estimates of $\beta$. For details on the iterative algorithm for estimation of $\alpha$, please refer to \cite{gee4}. Please note that it is one of the strengths of GEE that even if the chosen correlation structure is not accurate, the model still gives consistent results.

\section{Data availability}
To protect study participants’ privacy and consent, collected data will not be publicly available. Since the participants did not consent to share their data to the third party researchers, the datasets analyzed in this work are not publicly available. However, the code for models and some pre-processed features can be provided by the corresponding author on reasonable request.

\pagebreak

\maketitle
\section{Supplementary Information}
\subsection{Additional Results}

\begin{figure}[!htb]
	\centering
	\includegraphics[scale=0.55]{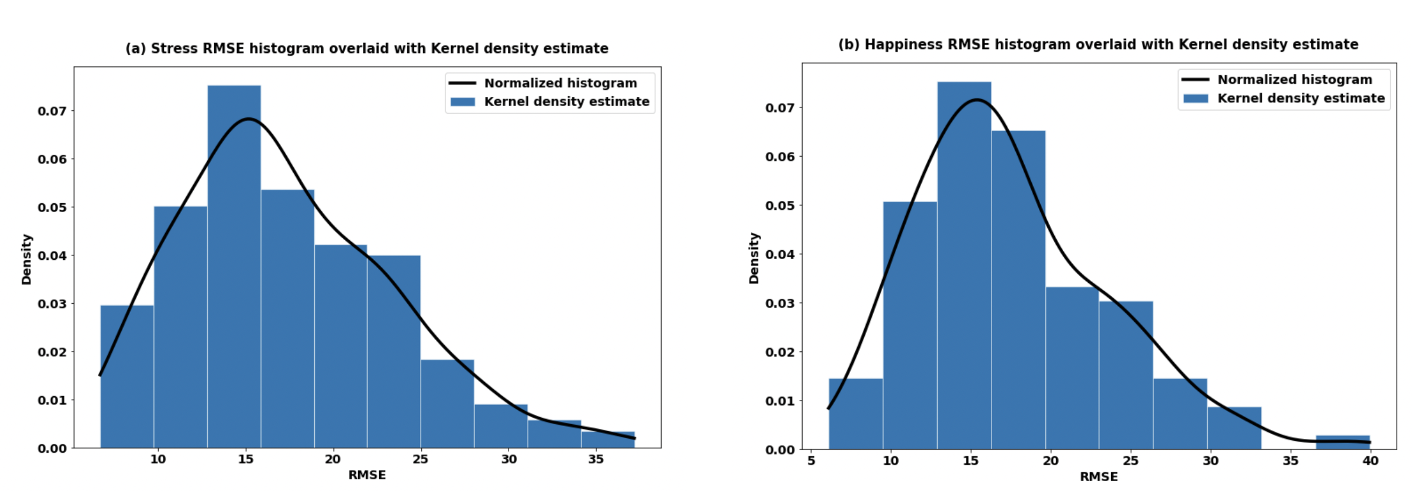}
	\caption{(S) Distribution of the RMSE. Normalized histogram is shown in blue bars and  a kernel density function is fitted to it.}
	\label{fig:rmse}
\end{figure}

\begin{table}[!htb]
\begin{tabular}{|l|ll|ll|}
\hline
\multirow{2}{*}{}                & \multicolumn{2}{l|}{Average stress score} & \multicolumn{2}{l|}{Standard deviation in stress score} \\ \cline{2-5} 
                                 & \multicolumn{1}{l|}{Coefficient}    & P-Value   & \multicolumn{1}{l|}{Coefficient}     & P\_Value    \\ \hline
Eigenvalue centrality            & \multicolumn{1}{l|}{-9}            & 0.004     & \multicolumn{1}{l|}{3.5}             & 0.01        \\ \hline
Small Degree (D\textless{}4)     & \multicolumn{1}{l|}{0.5}           & 0.8      & \multicolumn{1}{l|}{-1.2}            & 0.13        \\ \hline
Large Degree (D\textgreater{}4 ) & \multicolumn{1}{l|}{-2}           & 0.4      & \multicolumn{1}{l|}{-2}            & 0.17        \\ \hline
Closeness Centrality & \multicolumn{1}{l|}{1.5}           & 0.8     & \multicolumn{1}{l|}{-0.3}            & 0.9        \\ 
\hline
Pagerank Centrality & \multicolumn{1}{l|}{0.01}           & 0.8      & \multicolumn{1}{l|}{-0.01}            & 0.49        \\ 
\hline
\end{tabular}
\caption{(S) GEE results for model between true happiness score and graph centrality metrics.}
\label{tab:true_happy}
\end{table}

\begin{figure}[!htb]
    \centering
    \subfloat[][\centering Smaller graph with 10 nodes]{\includegraphics[width=8 cm]{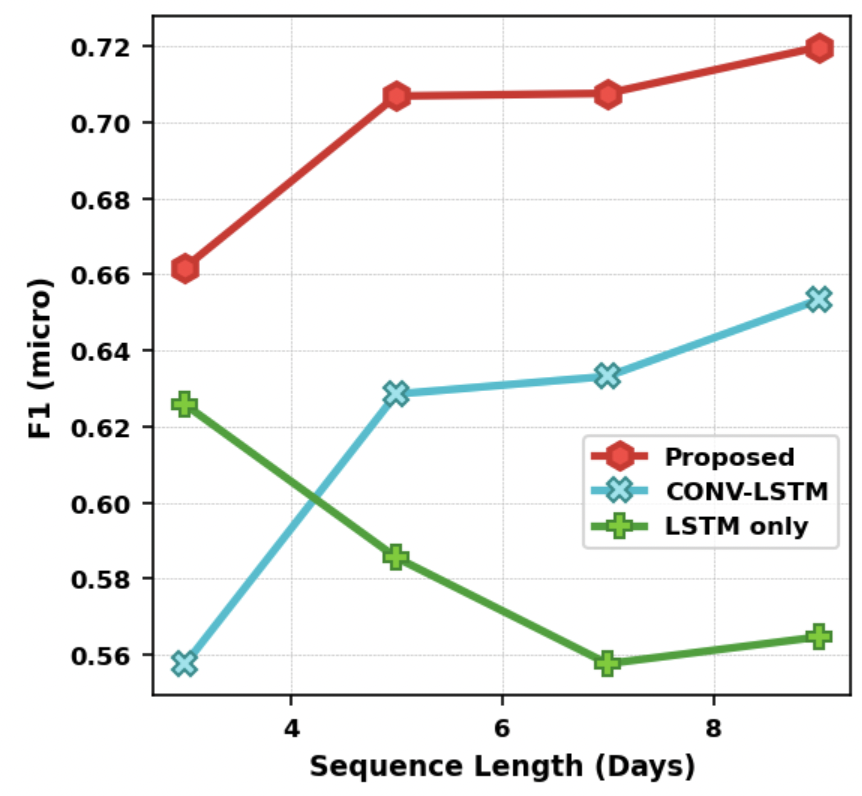} }%
    \quad
    \subfloat[][\centering Larger graph with 15 nodes]{\includegraphics[width=8cm]{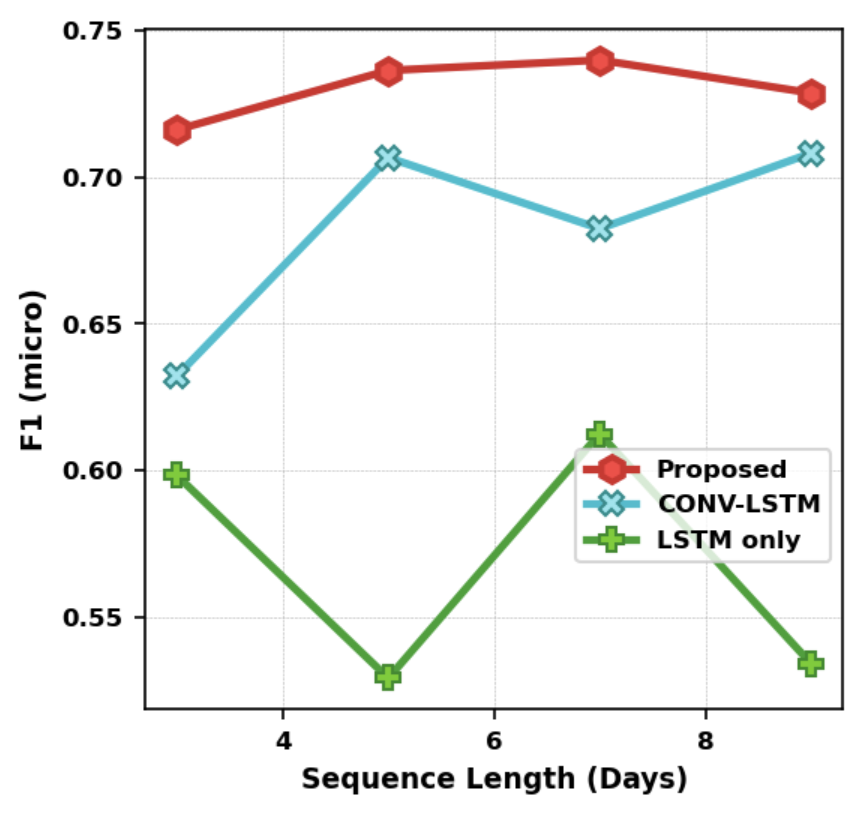} }%
    \caption{(S) Impact of temporal memory on happiness prediction}%
    \label{fig:seq_happiness}%
\end{figure}

\begin{figure}[h]
    \centering
    \subfloat[][\centering Impact of graph size on stress RMSE]{\includegraphics[width=7.5 cm]{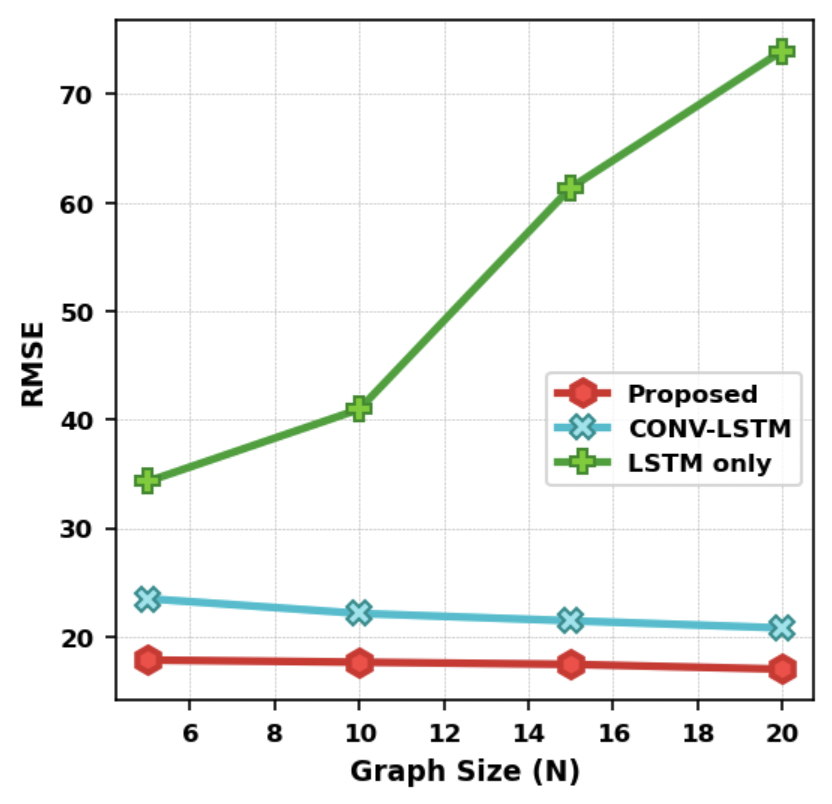} }%
    \qquad
    \subfloat[][\centering Impact of graph size on happiness RMSE]{\includegraphics[width=7.5cm]{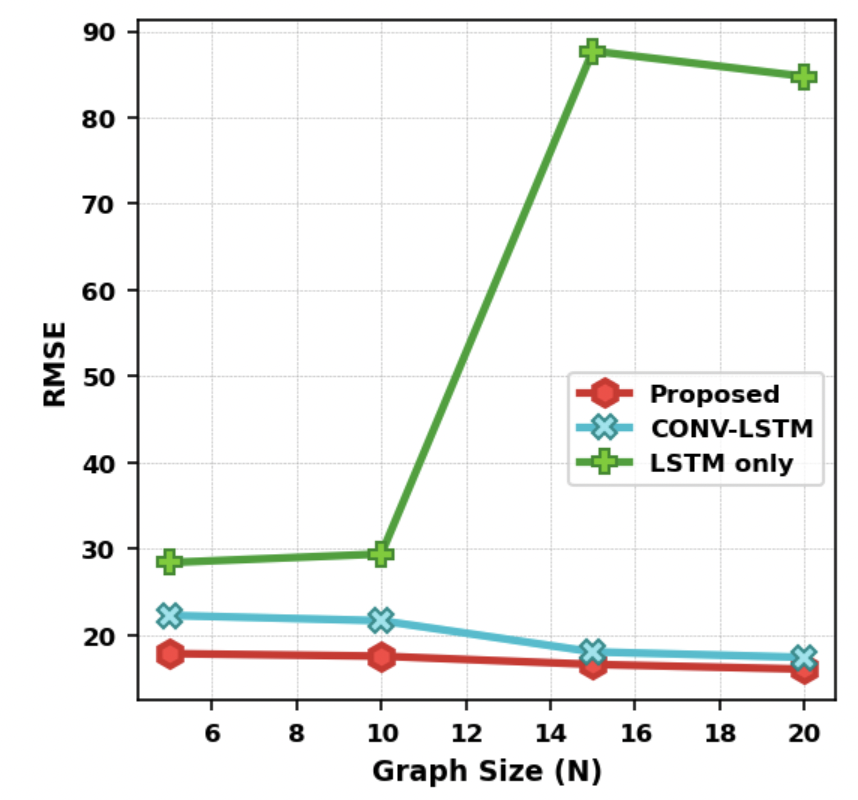} }%
    \caption{(S) Impact of graph size on root mean squared error}%
    \label{fig:happy}%
\end{figure}

\subsection{ Graph Convolutional Networks}
In this section, we provide a brief derivation of GCN forward propagation rule which incorporated the graph degree and laplacian matrix.
 The normalized graph Laplacian $L$ is a symmetric matrix defined as,
\begin{equation*}
	L = I_N - D^{-1/2}AD^{-1/2}
\end{equation*} 
where $I_N$ is $N\times N$ identity matrix.
 The spectral convolution on a graph is simply fourier domain multiplication of node feature signal $x$ (assume scalar for ease of understanding) with a filter $h_\theta$ parametrized by $\theta$\cite{GCN},
\begin{equation}
h(\theta) \ast x = U h_{\theta}(\Lambda)U^Tx
\end{equation}
where $U$ and $\Lambda$ are the eigenvector and eigenvalue matrix of $L$: $L = U\Lambda U^T$. To avoid the expensive computational cost of eigenvalue decomposition of L, a truncated Chebychev polynomial expansion of $h_\theta (\Lambda)$  is utilized to obtain the following simplified expression\cite{wavelet},
\begin{equation}\label{scaler}
	h(\theta) \ast x = L\theta x
\end{equation}
The details are skipped in the interest of space. For the detailed derivation, please refer to \cite{GCN}\cite{wavelet}. The expression in eq. \eqref{scaler} can be generalized for any signal $X\in \mathbb{R}^{NxC} $ for graph size $N$ and channels $C$,
\begin{equation}\label{conv}
	h(\theta) \ast x  = \tilde{D}^{-1/2} \tilde{A} \tilde{D}^{-1/2} X \Theta
\end{equation}
where $\tilde{A}=A+I_N$ is the adjacency matrix with added self-loops and $\tilde{D_{ii}}=\sum_{j}\tilde{A} _{ij}$. $\Theta$ represents the network parameters to be learned.

\bibliographystyle{ieeetr}
\bibliography{ref1.bib}%

\end{document}